\documentclass[12pt]{article}
\usepackage[T1]{fontenc}
\usepackage[utf8]{inputenc}
\usepackage[a4paper]{geometry}
\usepackage{float}
\usepackage{amsmath}
\usepackage{amssymb}
\usepackage{graphicx}
\usepackage{setspace}
\usepackage{esint}
\usepackage[unicode=true]
 {hyperref}

\makeatletter
\@ifundefined{date}{}{\date{}}

\usepackage{lmodern}
\usepackage[font=small,labelfont=bf]{caption}

\makeatother

\begin{document}
\title{A few paradoxes of Gauss' law\\
and how to avoid them}
\author{Marcin Kościelecki$^{1}$, Piotr Nieżurawski$^{2}$ \\
 {\small{}{}$^{1}$Department of Mathematical Methods in Physics,
Faculty of Physics, University~of~Warsaw} \\
 {\small{}{}ul.~Pasteura~5, 02-093 Warsaw, Poland} \\
 {\small{}{}$^{2}$Physics Education Laboratory, Faculty of Physics,
University of Warsaw} \\
 {\small{}{}ul.~Pasteura~5, 02-093 Warsaw, Poland} \\
 \textit{\small{}{}marcin.koscielecki@fuw.edu.pl}{\small{}{}, }\textit{\small{}{}piotr.niezurawski@fuw.edu.pl}{\small{}{}} }
\maketitle
\begin{abstract}
We present a few charge distributions for which the application of
Gauss' law in its integral form, as typically outlined in standard
textbooks, results in a~contradiction.\textbf{ }We identify the root
cause of such contradictions and put forward a~solution to resolve
them.\textbf{}\\
\\
 {01.55.+b, 01.40.-d, 03.50.De, 41.20.Cv}
\end{abstract}
\begin{quotation}
\noindent \begin{flushright}
\begin{minipage}[t]{0.8\columnwidth}%
\begin{quotation}
\noindent \emph{- So, we're setting out to find whatever this key
unlocks?}

\noindent \emph{- No. If we don't have the key, we can't open whatever
we don't have that it unlocks. So what purpose would be served in
finding whatever need be unlocked, which we don't have, without first
having found the key what unlocks it?}

\noindent \emph{- So, we're setting out to find this key?}

\noindent \emph{- Now you're not making any sense at all.}
\begin{flushright}
Pirates of the Caribbean: Dead Man's Chest, 2006
\par\end{flushright}
\end{quotation}
\end{minipage}
\par\end{flushright}
\newpage{}
\end{quotation}

\section{Introduction}

When teaching classical electrodynamics, we quite often meet students
who begun their knowledge of electrostatics with questions concerning
'infinite, uniformly charged rod or cylinder' or 'infinite uniformly
charged plane'. Students are expected to calculate the electric field
resulting from these charge distributions. Usually the integral form
of Gauss' law is used (e.g.~\cite{Resnick}, \cite{Berkeley}). These
problems have their charm. Their results are obtained without complex
calculations through the use of the symmetry of the system. 

However, in our experience teaching Gauss' law in integral form has
unexpected side-effects. When we ask students to calculate the electric
flux through the surface of the sphere with the dipole source, we
usually get the right answer that the flux is equal to zero, because
the total charge in the sphere is equal to zero. Unfortunately, many
students add, even when not asked, that the electric field of a dipole
is also zero!

Another unintended effect of the use of an integral form of the Gauss'
law is the belief of some students that the Gauss' law is used to
find the electric field at a given point and the Gauss' law can always
be applied. It seems that most students -- and not only students
-- are convinced that a solution exists for any charge distribution.
The trouble is that \emph{``Gauss' law + symmetry''} techniques
are often unreliable when it comes to problems that are slightly modified
versions of the standard examples presented during lectures and in
textbooks.

We would like to present a few charge distributions that can help
to facilitate the discussion of limitations and applicability of the
Gauss' law, and of the electrostatics in general. Such examples should
be the follow-ups of classical problems like the 'infinite, uniformly
charged rod' or 'infinite uniformly charged plane'. It would be beneficial
if these examples were used to reflect on the following questions:
\begin{itemize}
\item Is it always possible to use the integral form of Gauss' law to find
the electric field even if symmetries are present?
\item Does the solution exist for any charge distribution?
\item How to check if the solution does exist?
\end{itemize}

\section{Paradoxes\label{sec:Paradoxes}}

In this article we use the following form of Gauss' law

\[
\varepsilon_{0}\ointop_{\partial V}\vec{E}\cdot d\vec{s}=\intop_{V}\rho\,dv
\]

where $\varepsilon_{0}$ is the vacuum permittivity, $\vec{E}$ is
the electric field, $V$ is a volume and $\partial V$ its surface,
$d\vec{s}$ is an infinitesimal area vector pointing to the outside
of the volume $V,$ and $\rho$ is the volume charge density.

Let us consider some paradoxes arising from the application of Gauss'
law and the assumptions about the symmetry of the electric field. 

\subsection{Infinite, uniformly charged world and one sphere}

We will begin with the simplest example -- 'the infinite, uniformly
charged world', that is, an infinite space in which the volume charge
density $\rho$ is constant.\footnote{Similar problem could be found for Newtonian gravity and is known
as a Seeliger's paradox. However Seeliger tried to modify the formula
for gravitational forces instead of considering some distributions
of matter as 'unphysical'. The birth of General Relativity reduced
the importance of this problem.} If one would like to find the reason for considering this distribution
not only as a theoretical curiosity, it might be naively suggested
that this is 'an approximation' of an uniformly charged,\textbf{ }large
cube and the results will be considered valid only near the center
of the cube to ignore the effects 'caused by' the boundary of the
cube. Similarly, a square plate of a capacitor is 'approximated' by
an infinite plane.

\begin{figure}
\begin{centering}
\includegraphics[height=7cm]{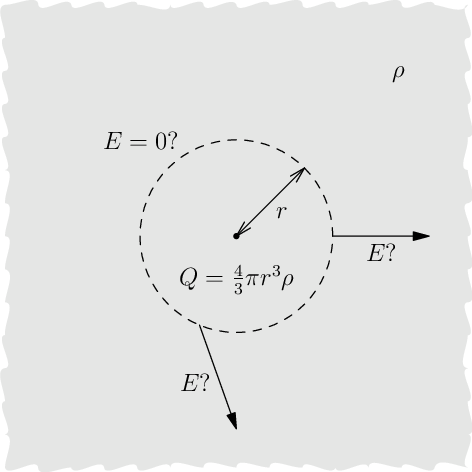} 
\par\end{centering}
\caption{\label{fig:sfera}The infinite world, uniformly charged with density
$\rho$. The grey area represents a two-dimen\-sional cross-section
of the fragment of the world. The dashed line represents the sphere
through which the electric flux is calculated, which according to
the Gauss' law should be equal to the charge $Q$ in the sphere. What
will be the electric field in such a world?}
\end{figure}

We choose a sphere of radius $r$ whose center is at a certain, fixed
point (Fig.~\ref{fig:sfera}). Now let us apply the integral form
of Gauss' law. Since the sphere has the charge 
\[
Q=\intop_{V}\rho\,dv=\frac{4}{3}\pi r^{3}\rho
\]
the flux of the electric field $\vec{E}$ through the sphere may not
be equal to 0. This leads to the conclusion that the field $\vec{E}$
cannot be equal to 0 as well. On the other hand, the considered world
is homogeneous, there are no distinguished points and directions,
so according to usual symmetry arguments, $\vec{E}$ must be zero
everywhere.

Two ways to determine the electric field lead to two conflicting conclusions.
In our opinion the conclusion is one: the problem is ill-defined (see
section \ref{sec:The-source-of-problems}).

\subsection{Infinite, uniformly charged world and two spheres}

\begin{figure}[H]
\begin{centering}
\includegraphics[height=7cm]{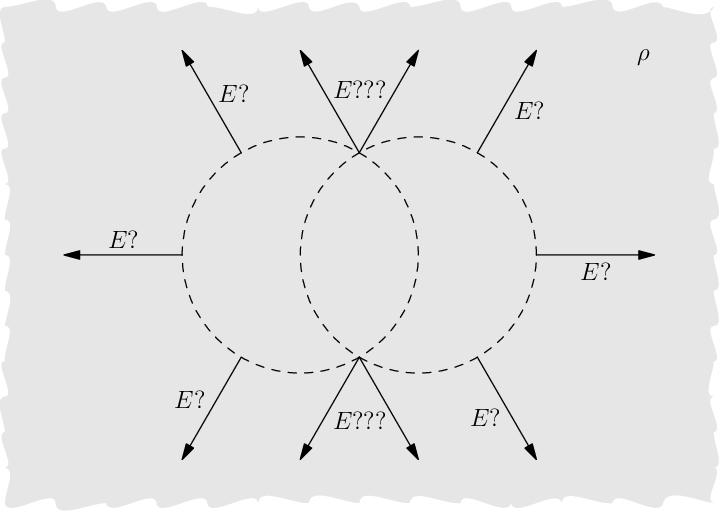} 
\par\end{centering}
\caption{\label{fig:sfera-arrows-E}The infinite world, uniformly charged with
density $\rho$ (a two-dimen\-sion\-al cross-section). The dashed
line represents the spheres through which the electric flux is calculated.
Assuming that the electric field has a spherical symmetry on the left
sphere, the electric field can be determined according to Gauss' law.
However, choosing another sphere one can obtain a different electric
field at the same point.}
\end{figure}

Yet another contradiction can be found in the infinite, uniformly
charged world. Let us try to determine the flux of the field $\vec{E}$
through the sphere centered at the selected point. According to Gauss'
law the field cannot be equal to zero on the entire sphere as the
flux is different from zero. Assuming that the field is directed radially
and has a fixed magnitude on the sphere -- this is suggested by the
symmetry of the problem -- we determine the electric field. Now,
we choose a center of the second sphere at another point, but let
this sphere pass through the point where the field $\vec{E}$ was
already been 'determined'. Of course, using the second sphere we get
a different vector field at the same point than with the first choice
of the sphere (Fig.~\ref{fig:sfera-arrows-E}). 

Again we obtained two contradictory results.

\begin{figure}
\begin{centering}
\includegraphics[height=7cm]{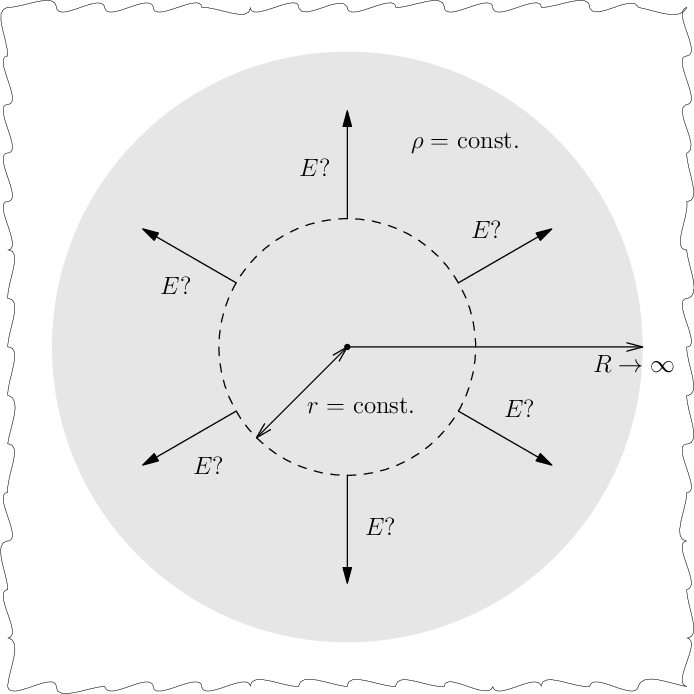} 
\par\end{centering}
\caption{\label{fig:sfera-arrows-expanding-ball}A ball of radius $R$, uniformly
charged with density $\rho$ (a two-dimensional cross-section is represented
by the grey area). The ball initially has a finite radius $R$. The
dashed line represents the sphere of radius $r$ through which the
electric flux is calculated. We want to know whether the electric
field remains unchanged at a fixed distance $r$ from the center when
$R\rightarrow\infty$ while the charge density $\rho$ is constant.
Then the whole infinite 3-dimensional space (represented by an irregular
box) will be uniformly charged.}
\end{figure}

\subsection{Expanding charged ball\label{subsec:Expanding-charged-ball}}

A similar paradox is found with a slightly different approach. Let
us consider a uniformly charged ball of radius $R$ in the empty space.
We know that inside the ball the field $\vec{E}$ has a radial direction
and the magnitude of 
\[
E=\frac{\rho r}{3\epsilon_{0}}
\]
where $r$ is the distance from the center of the ball (Fig.~\ref{fig:sfera-arrows-expanding-ball}).
This result is obtained by using the spherical symmetry of the problem
and applying the integral form of Gauss' law. In the following we
keep $r$ and charge density $\rho$ constant. The paradox arises
when the radius of the ball reaches limit $R\rightarrow\infty$. The
previously obtained value of the field $\vec{E}$ is different from
zero and independent of $R$. However, it seems that the magnitude
of the field should approach zero with $R\rightarrow\infty$ because
after this 'swelling' of the ball we get an infinite, uniformly charged
space and everywhere $E$ should be equal to 0 according to symmetry
arguments, as in the first example. This reasoning can be repeated
for a uniformly charged cylinder of infinite height and for an infinite
plate. In both cases, the field $\vec{E}$ measured at a point lying
inside the cylinder or the plate does not depend on the size of these
charged objects. Enlarging them so that they occupy infinite space,
we get a similar paradox. 

When we expand a charged sphere to fill the entire world, we maintain
a nonzero electric field -- even though the symmetry of the world
suggests it should be zero. Moreover, by choosing a different point
as the center of the sphere or another uniformly charged, expanding
body, we will obtain a different field at the same point.

\subsection{Infinite sandwich}

\begin{figure}[h]
\begin{centering}
\includegraphics[width=1\textwidth]{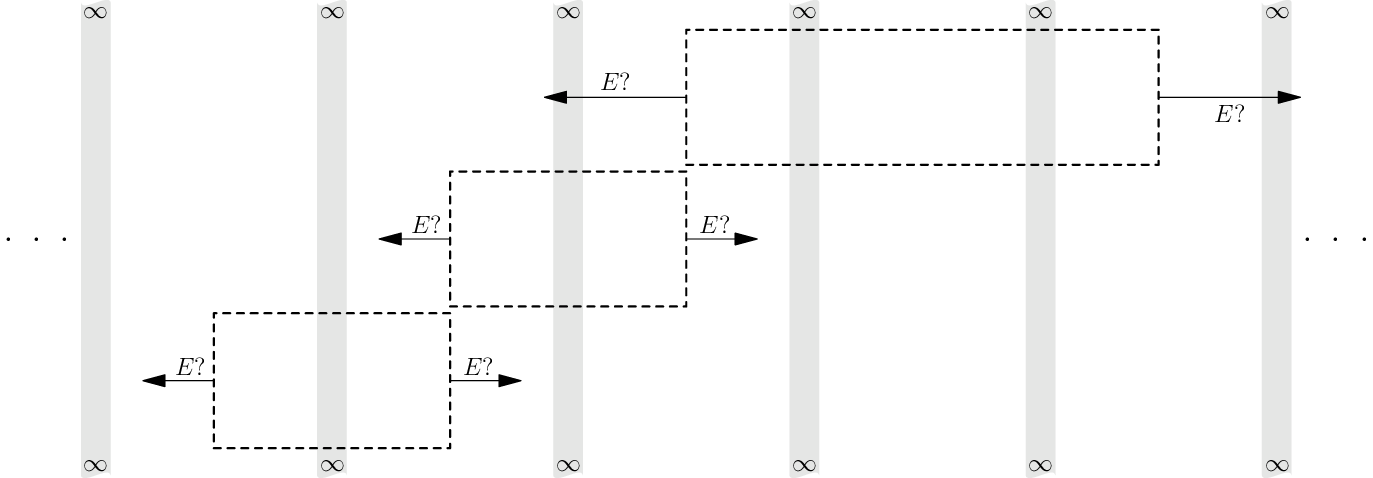} 
\par\end{centering}
\caption{\label{fig:infinite-sandwich-one-enclosed} 'An infinite sandwich'
-- an infinite number of equally spaced, parallel, uniformly charged
infinite planes or plates (two-dimensional cross-section). The dashed
line represents a rectangular box through which the electric flux
is calculated. What is the electric field? }
\end{figure}

Another paradox arises when considering 'an infinite sandwich'. Let
us imagine a world with an infinite number of parallel and equally
distant infinite planes or plates. Each plate is charged with the
same constant surface density. If we follow the methods used for a
single plate, it turns out that the field in 'the infinite sandwich'
depends on how many and which plates are included in the rectangular
box that is used to calculate the flux and the charge in Gauss' law
(Fig.~\ref{fig:infinite-sandwich-one-enclosed}). For example, if
we double the charge contained in the box, according to Gauss' laws
the magnitude of the electric field on the sides of the box, which
are parallel to the plates, should double. At a given point we can
obtain different fields, depending on the choice of the box. Again,
we obtain a contradiction.

\subsection{Recapitulation}

It can be concluded that if one uses an approach in which the finite,
computationally difficult distributions are replaced by simple, infinite
distributions, serious problems may arise. 'An infinite sandwich'
could be naively used as a convenient model of a system consisting
of thousands square, charged plates arranged in parallel at equal
distances. It is clear, however, that in this case 'the stretching
to infinity' or applying periodic boundary conditions lead to serious
problems.

\section{The source of problems\label{sec:The-source-of-problems}}

\begin{figure}[h]
\begin{centering}
\includegraphics[width=1\textwidth]{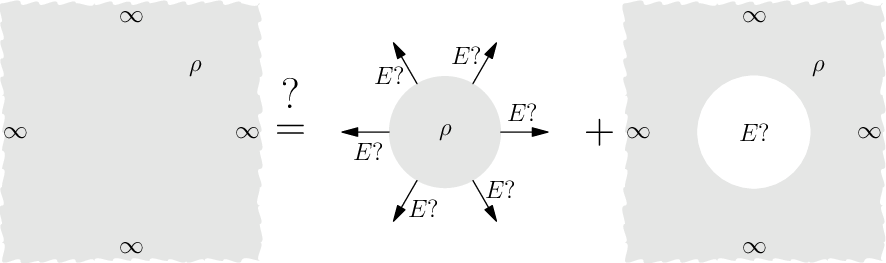} 
\par\end{centering}
\caption{\label{fig: sphere-is-superposition-valid} Can one always neglect
charges placed outside the Gaussian surface? Can a universe filled
with uniform charge density be split into a compact ball and its complement?}
\end{figure}

Examples from the previous paragraph illustrate that in order to solve
problems in a 'Gauss way' the following assumptions were made: 
\begin{enumerate}
\item about the existence of solutions, 
\item about the symmetry of potential solutions, 
\item about the automatic consistency of any solution with boundary conditions, 
\item that the integral version of Gauss' law can be used in any case (i.e.~in
the noncompact support case), 
\item that the total charge in the universe does not have to be finite, 
\item that using Gauss' law one can always ignore charges lying outside
the surface of integration (Fig.~\ref{fig: sphere-is-superposition-valid}).
\end{enumerate}

Students often are not conscious of the above assumptions. Therefore,
the paradoxes presented in section \ref{sec:Paradoxes} could be of
didactic benefit. 

\section{Resolution}

If two different conclusions can be obtained for the given charge
distribution, then there is no unique solution for this case. As we
show in \cite{Koscielecki-Niezurawski-Infinite-plate}, the infinite
plate is the more subtle example of such charge distribution for which
direct calculation shows that the unique electric field does not exist.
However, there exist infinite charge distributions for which direct
calculation leads to the unique electric field. The simple example
is the infinite wire \cite{Koscielecki-Niezurawski-Infinite-plate}.
Therefore, each infinite distribution should be treated separately. 

We recommend to use the integral form of Gauss' law\textbf{ }as the
tool\textbf{ }that allows to find solutions quickly if it is known,
that the solution exists. Example:\textbf{ }It is known that the solution
for an infinite wire exists \cite{Koscielecki-Niezurawski-Infinite-plate},
thus we can calculate the electric field using the Gauss law.

Alternatively, the Gauss' law can be used to show that there are at
least two different conclusions regarding the electric field, thus
the solution does not exist. Therefore the lack of solution can be
shown by using the Gauss' law. Example: In the infinite uniformly
charged universe for different Gaussian surfaces we can obtain different
values and directions of the electric field in the given point, therefore
there is no unique solution -- the problem is ill-defined.

It can be argued that the results obtained for infinite spatial distributions
-- whether they exist or not -- are a good approximation of the
results that we would expect to find for finite spatial distributions.
Thus, an infinite, uniformly charged rod is a good -- and convenient
-- approximation of a long, finite rod, and an infinite, uniformly
charged plane is an approximation of a large, finite plate of a capacitor.
Unfortunately, students rarely have a chance to find out how good
these approximations are. For example, once in the lifetime a student
should calculate the field due to a finite, rectangular, uniformly
charged plate and see the difference between the result and $\sigma/(2\varepsilon_{0})$
\cite{Koscielecki-Niezurawski-Infinite-plate}. 

The issue is fundamental and should result in the conclusion that
while applying laws of physics one should also look for the underlying
assumptions -- such a proposal seems as obvious as it is often overlooked.

\section{Summary}

We present several charge distributions for which determination of
the electric field by means of the integral Gauss' law and standard
methods used in textbook problems leads to a contradiction. We show
that the Gauss' law can be used to falsify statements about existence
of solution, for example in the infinite uniformly charged world.
As the final check if there the electric field exists we suggest to
use a Coulomb integral without symmetry assumptions \cite{Koscielecki-Niezurawski-Infinite-plate}.
If it is known, that the solution exists, then the Gauss' law can
be used to quickly find solutions.

It is our assertion that when an infinite total charge is given and
the electric field should be found, there are only the following important
cases:

1. The problem has a solution if and only if there is a Coulomb integral.

\emph{More detailed description:} For a given charge distribution
we calculate the Coulomb integral. If this integral exists (without
any assumptions about symmetry), then the problem has a solution.
For example, an infinite uniformly charged straight wire.

2. If the problem has a solution, then we can obtain it using the
Gauss' law or Coulomb integral using symmetry.

\emph{More detailed description:} For an uniformly charged infinite
straight wire we know that the Coulomb integral exists, so we can
use symmetry to get the result quickly and easily. We cannot use this
method to prove that a solution exists.

3. If using the Gauss' law and symmetry we come to a contradiction
(at least two contradictory solutions), then the problem has no solution. 

\emph{More detailed description:} For an infinite sandwich or an infinite
uniformly charged world we can obtain from Gauss' law, assuming symmetry
of the electric field, different fields at a given point. So the solution
does not exist.

It appears that a verification of the solution's existence when addressing
electrostatic problems could prove advantageous in the educational
process, as it enables the avoidance of the aforementioned paradoxes.
Otherwise it is not clear for students why in the case of some charge
distributions with infinite total charge the integral Gauss' law leads
to correct solutions but in other cases does not. Moreover, the presented
paradoxes alone can be used to activate students, to trigger creative
anxiety which can lead to a general discussion about the conditions
of applicability of the laws of physics. For example students may
formulate similar paradoxes in magnetostatics for Ampere's law.

The authors would like to thank Andrzej Majhofer, Kazimierz Napiórkowski,
and~Robin \& Tad Krauze for fruitful discussions and valuable comments.

\section*{Answers to the frequently asked questions}

We have encountered comments expressing disbelief that the standard
use of Gauss' law can lead to errors in certain cases. In this spirit
of constructive engagement, we would like to address the most commonly
raised concerns in a forthright yet respectful manner:

\bigskip{}

Statement: \emph{There are no paradoxes -- just choose the right
surface and symmetry.}

Answer: This is not true. Gauss' law can be applied to various surfaces,
there are no \textquotedbl only right\textquotedbl{} ones. Moreover,
using the Gauss' law and the fixed surface one looses information
about the electric field component that is tangent to this surface.

\bigskip{}

Statement: \emph{Gauss' law cannot be applied in this (e.g.~uniformly
charged world) case.}

Answer: Yes and no. In this case the Gauss' law can be used to show
that the solution does not exist.

\bigskip{}

Statement: \emph{The calculation using Gauss' law for infinite plate
is very easy but what you do in \cite{Koscielecki-Niezurawski-Infinite-plate}
is much more complicated.}

or 

\emph{There are such mathematical details that (even if correct) I
would not prefer to see in a typical physics textbook for students.}

Answer: The statement that there is no solution for infinite plate
is as deep as the difference between 
\begin{align*}
\lim_{a\rightarrow\infty,\,b\rightarrow\infty}(a-b)
\end{align*}

and
\begin{align*}
\lim_{a\rightarrow\infty}(a-a)
\end{align*}

If you can explain difference between these limits, then you can explain
why electric field due to an uniformly charged infinite plate is indefinite.

\bigskip{}

Statement: \emph{But I guess we also have to consider the charge outside
the contour in an uniformly charged world case?}

Answer: No, in Gauss' law only the charge inside the surface, in the
finite volume, is important. However, as we show, Gauss' law has a
limited range of applicability.

\bigskip{}

Statement: \emph{Infinite objects are a convenient approximation,
so why not use infinite plates, capacitors and worlds?}

Answer: First disadvantage --- lack of information on how good this
approximation is; second disadvantage --- the boundary transition
from a finite object to an infinite object sometimes leads to the
conclusion that the electric field does not exist (example: an expanding
ball in section \ref{subsec:Expanding-charged-ball} in this article
or article \cite{Koscielecki-Niezurawski-Infinite-plate}).

\bigskip{}

Statement: \emph{Choose a method that gives a good result, and assume
that other methods or results are nonphysical.}

Answer: This is undoubtedly a quick method of avoiding contradictions...
but does not convince everyone.
\end{document}